\newenvironment{TH4.1}{\vspace{6pt}\noindent{\bf Proof of Theorem
4.1:}} { \rule{0.075in}{0.075in} \vspace{11pt}}
\newenvironment{TH4.2}{\vspace{6pt}\noindent{\bf Proof of Theorem
4.2:}} { \rule{0.075in}{0.075in} \vspace{11pt}}

\documentclass[10pt,twoside]{article}
\usepackage{ndst}

\begin{document}
\setcounter{page}{1}

\newarticle

\label{Sample_NDST: FirstPage}

\Vykh{Sample_NDST: 1}{Sample_NDST: 10}

\renewcommand{\ehkol}{E.V. Krishnan, M. Al Ghabshi and M. Alquran}
\renewcommand{\ohkol}{$(G'/G)$-expansion method and Weierstrass elliptic function method,
{\bf 19}\,(2) (2019)
\pageref{Sample_NDST:1}--\pageref{Sample_NDST:10}}

\title{$(G'/G)$-Expansion Method and Weierstrass Elliptic Function Method Applied to Coupled Wave Equation}  
\author{E.V. Krishnan\,$^{1,*}$, M. Al Ghabshi\,$^{1}$, M. Alquran\,$^{2}$} 
\address{$^1$ Department of Mathematics, Sultan Qaboos University, PO Box 36, Al Khod 123, Muscat, Sultanate of Oman\\
$^2$ Department of Mathematics and Statistics, Jordan University of
Science and Technology, P.O.Box (3030), Irbid (22110), Jordan}

\renewcommand{\thefootnote}{} \footnote{$^*$ Corresponding author:
\url{krish@squ.edu.om} and \url{drkrish52@gmail.com}}

\date{Received: December 5, 2018;\ \ \ \ Revised: October 9, 2019}

\abstract{This paper deals with the exact solutions of a nonlinear
coupled coupled wave equation. The $(G'/G)$-expansion method has
been applied to derive kink solutions and singular wave solutions.
The restrictions on the coefficients of the governing equations have
also been investigated. Solitary wave solutions have also been
derived for this system of equations using
Weierstrass elliptic function method.}  



\bigskip

\section{Introduction}
Nonlinear evolution equations (NLEEs) govern several physical
phenomena which appear in various branches of science and
engineering \cite{R1, R2, R3, R4, R5}. Exact solutions of NLEEs shed
more light on the various aspects of the problem, which, in turn,
leads to the applications. Several methods such as the tanh method
\cite{R6, R7, R8, R9, MAR5, MAR6}, exponential function method
\cite{R10}, Jacobi elliptic function (JEF) method \cite{R11, MAR1,
MAR2, extra}, mapping methods \cite{R12, R13, R14, R15, R16, R17},
Hirota bilinear method \cite{jaradat1, jaradat2} and
trigonometric-hyperbolic function methods \cite{trig1, trig2, trig3}
have been applied in the last few decades and the results have been
reported. Also, many physical phenomena have been governed by
systems of partial differential equations (PDEs) and there have been
significant contributions in this area \cite{R18, R19}.\\
\\
In this paper, we use the $(G'/G)$-expansion method \cite{R20, R21,
R22, MAR3, MAR4} to find some exact solutions for a nonlinear
coupled wave equation \cite{R23}. The paper is organized as follows.
In Section 2, we give a mathematical analysis of the
$(G'/G)$-expansion method, in Section 3, we find kink solutions and
singular wave solutions of the nonlinear coupled wave equation, in
Section 4, we use the Weierstrass elliptic function (WEF) method
\cite{R24} to derive SWSs of the system of equations, in Section 5
we write down the conclusion.

\section{The $(G'/G)$-Expansion Method}

Consider the nonlinear partial differential equation (PDE)
\begin{eqnarray}
P(u, u_t, u_x, u_{tt}, u_{xt}, u_{xx},...) =0,
\end{eqnarray}
where $u(x,t)$ is an unknown function, $P$ is a polynomial in $u =
u(x,t)$ and its various partial derivatives. The traveling wave
variable $\xi = x - ct$ reduces the PDE (1) to the ordinary
differential equation (ODE)
\begin{eqnarray}
P(u, -c u^{\prime}, u^{\prime}, -c^2 u^{\prime\prime}, -c
u^{\prime\prime}, u^{\prime\prime},...) = o,
\end{eqnarray}
where $u = u(\xi)$ and $^{\prime}$ denotes differentiation with
respect to $\xi$.\\
\\
We suppose that the solution of equation (2) can be expressed by a
polynomial in $\displaystyle \left(\frac{G^{\prime}}{G}\right)$ as
follows:

\begin{eqnarray}
\displaystyle u(\xi)= \sum_{i=0}^m a_i
\left(\frac{G^{\prime}}{G}\right)^i, \,\,\,a_m \neq 0,
\end{eqnarray}
where $a_i (i = 0,1,2,..)$ are constants.\\
\\
Here, $G$ satisfies the second order linear ODE
\begin{eqnarray}
\displaystyle G^{\prime\prime}(\xi) + \lambda G^{\prime} (\xi) + \mu
G(\xi) = 0,
\end{eqnarray}
with $\lambda$ and $\mu$ being constants. The positive integer $m$
can be determined by a balance between the highest order derivative
term and the nonlinear term appearing in equation (2). By
substituting equation (3) into equation (2) and using equation (4),
we get a polynomial in $\displaystyle G^{\prime}/G$. The
coefficients of various powers of $\displaystyle G^{\prime}/G$ give
rise to a set of algebraic
equations for $a_i \,(i = 0,1,2, ..., m),\,\lambda$ and $\mu$.\\
\\
The general solution of equation (4) is a linear combination of
${\rm{sinh}}$ and ${\rm{\cosh}}$ or of sine and cosine functions if
$\Delta = \lambda^2 - 4 \mu > 0$ or $\Delta = \lambda^2 - 4 \mu <
0$, respectively. In this paper we consider only the first case and
so,
\begin{eqnarray}
G(\xi) = \displaystyle e^{-\lambda \xi/2}
\left(C_1\,{\rm{sinh}}(\frac{\sqrt{\lambda^2 - 4 \mu}}{2}\,\xi) +
C_2\,{\rm{cosh}}(\frac{\sqrt{\lambda^2 - 4 \mu}}{2}\,\xi)\right),
\end{eqnarray}
where $C_1$ and $C_2$ are arbitrary constants.

\section{A Coupled Wave Equation}

Consider the system of PDEs
\begin{equation}
 \displaystyle u_{t}+ \alpha v^2 v_{x}+ \beta u^2 u_{x} + \eta u u_{x}+ \gamma u_{xxx}=0,
\end{equation}
\begin{equation}
 \displaystyle v_t+ \sigma(uv)_x+ \epsilon v v_x=0,
\end{equation}
where, $\alpha, \beta, \eta, \gamma, \sigma$ and $\epsilon$ are
constants.
\\
\\
We seek TWSs of equations (6) and (7) in the form $ u = u(\xi),\quad
v = v (\xi) ,\quad \xi = x- c t.  $
\\
\\
Then equations (6) and (7) give\\
\begin{equation}
\displaystyle - c u^{\prime} + \alpha v^2v^{\prime} + \beta u^2
u^{\prime} + \eta u u^{\prime}+ \gamma u^{\prime\prime\prime} = 0,
\end{equation}
\begin{equation}
\displaystyle -c v^{\prime} + \sigma (u v)^{\prime}+ \epsilon v
v^{\prime} = 0.
\end{equation}
Integrate equation (9) with respect to $ \xi $
\begin{equation}
-c v + \sigma (u v) + \dfrac {\epsilon}{2} v ^2= k,
\end{equation}
where $k$ is the integration constant.
\\
\\
Dividing equation (10) by $v$, we obtain
\begin{equation}
- c + \sigma u + \displaystyle \frac{\epsilon}{2} v = \frac{k}{v}.
\end{equation}
\\
\\
So, for the solutions to be uniformly valid, the integration
constant $k$ should be set equal to $0$.
\\
\\
Therefore, equation (11) can be written as
\begin{eqnarray}
v = \displaystyle \frac{2 (c - \sigma u)}{\epsilon}.
\end{eqnarray}
\\
\\
Substituting equation (12) into equation (8), we obtain
\begin{equation}
-c u^{\prime} -\dfrac{8 \alpha \sigma }{\epsilon^ 3}(c^2 -2 c \sigma
u + \sigma ^2 u^2 )u^{\prime}+ \beta u^2 u^{\prime} + \eta u
u^{\prime}+ \gamma u^{\prime\prime\prime} = 0.
\end{equation}
\\
\\
Integrating equation (13) with respect to $ \xi $ and assuming the
boundary conditions $\displaystyle  u, u^{\prime}, u^{\prime\prime}
\longrightarrow 0 $\,\, $ {\rm{as}} \,\, \vert \xi \vert
\longrightarrow \infty $, \,\, we have
\begin{equation}
\gamma u^{\prime\prime} - \bigg \lbrace c +\dfrac{8 \alpha \sigma
c^2 }{\epsilon^ 3} \bigg \rbrace u + \bigg \lbrace \dfrac{\eta}{2} +
\dfrac{8 \alpha \sigma ^ 2 c }{\epsilon^ 3} \bigg \rbrace u^2 +
\bigg \lbrace \dfrac{\beta}{3} - \dfrac{8 \alpha \sigma^3 }{3
\epsilon^ 3} \bigg \rbrace u^3 = 0.
\end{equation}
\\
\\
We rewrite equation (14) as
\\
\\
\begin{equation}
\gamma u^{\prime\prime} + A u + B u ^2 +Cu^3 =0,
\end{equation}
\\
\\
where
\begin{eqnarray}
 A= - \bigg \lbrace c +\dfrac{8 \alpha \sigma c^2 }{\epsilon^ 3} \bigg \rbrace, \,\,\, B=\bigg \lbrace \dfrac{\eta}{2} + \dfrac{8 \alpha \sigma ^ 2 c }{\epsilon^ 3} \bigg \rbrace ,\; C=\bigg \lbrace \dfrac{\beta}{3} - \dfrac{8 \alpha \sigma^3 }{3 \epsilon^ 3} \bigg \rbrace.
\end{eqnarray}

With the change of variable $w = u + \delta$, equation (15) can be
reduced to

\begin{equation}
\gamma w^{\prime\prime} + c_1 w + c_2 w^3 + c_3 =0,
\end{equation}
where
\begin{eqnarray}
\displaystyle \delta = -\frac{B}{3 C},\,\,\,\,c_1 = \frac{3AC -
B^2}{3 C},\,\,\,\,c_2 = C,\,\,\,\,c_3 = \frac{2 B^3 - 9 ABC}{27
C^2}.
\end{eqnarray}
Assuming the expansion $\displaystyle w(\xi)= \sum_{i=0}^m a_i
\left(\frac{G^{\prime}}{G}\right)^i, \,\,\,a_m \neq 0$ in eq. (17)
and balancing the nonlinear term and the derivative term, we get $m
+ 2 = 3m$ so that $m = 1$.
\\
\\
So, we assume the solution of equation (17) in the form
\begin{eqnarray}
w(\xi) = a_0 + a_1 \displaystyle
\left(\frac{G^{\prime}}{G}\right),\,\,a_1 \neq 0.
\end{eqnarray}
So, we can obtain
\begin{eqnarray}
\displaystyle w^{\prime}(\xi) = - a_1
\left(\frac{G^{\prime}}{G}\right)^2 - \lambda a_1
\left(\frac{G^{\prime}}{G}\right) - \mu a_1,
\end{eqnarray}
\begin{eqnarray}
\displaystyle w^{\prime\prime}(\xi) = 2 a_1
\left(\frac{G^{\prime}}{G}\right)^3 + 3 a_1 \lambda
\left(\frac{G^{\prime}}{G}\right)^2 + (a_1 \lambda^2 + 2 a_1 \mu)
\left(\frac{G^{\prime}}{G}\right) + a_1 \lambda \mu,
\end{eqnarray}
\begin{eqnarray}
\displaystyle w^3(\xi) = a_1^3 \left(\frac{G^{\prime}}{G}\right)^3 +
3 a_0 a_1^2 \left(\frac{G^{\prime}}{G}\right)^2 + 3 a_0^2 a_1
\left(\frac{G^{\prime}}{G}\right) + a_0^3.
\end{eqnarray}
Now, substituting equations (19), (21) and (22) into equation (17)
and collecting the coefficients of $\displaystyle
\left(\frac{G^{\prime}}{G}\right)^i, i = 0,1,2,3$, we get
\begin{eqnarray}
\gamma a_1 \lambda \mu + c_1 a_0 + c_2 a_0^3 + c_3 = 0,
\end{eqnarray}
\begin{eqnarray}
\gamma a_1 \lambda^2 + 2 \gamma a_1 \mu + c_1 a_1 + 3 c_2 a_0^2 a_1
= 0,
\end{eqnarray}
\begin{eqnarray}
3 a_1 \lambda \gamma + 3 a_0 a_1^2 c_2 = 0,
\end{eqnarray}
\begin{eqnarray}
2 \gamma a_1 + c_2 a_1^3 = 0.
\end{eqnarray}
From equation (26), we get
\begin{eqnarray}
\displaystyle a_1 = \pm \sqrt{- \frac{2 \gamma}{c_2}}.
\end{eqnarray}
Equation (25) leads to
\begin{eqnarray}
\displaystyle a_0 = \pm \frac{1}{2} \lambda \sqrt{- \frac{2
\gamma}{c_2}}.
\end{eqnarray}
When $\mu = 0$ in equation (24), we get $\lambda = \displaystyle \pm
\sqrt{\frac{2 c_1}{\gamma}}$ and when $\lambda = 0$, \, we get  $
\displaystyle \mu = - \frac{c_1}{2 \gamma}.$ In both cases,
$\displaystyle \Delta = \lambda^2 - 4 \mu = \frac{2 c_1}{\gamma}.$
\\
\\
Equation (23) gives a constraint condition on the coefficients in
the governing equation.
\\
\\
\textbf{Case 1}:\,\,\,\,$\displaystyle \mu = 0,\,\,\lambda =
\sqrt{\frac{2 c_1}{\gamma}}.$

\begin{eqnarray}
u_1(x,t) = \displaystyle \pm \sqrt{- \frac{c_1}{c_2}}\left[1 +
\frac{(C_1 - C_2)\left(1 - {\rm{tanh}}\frac{1}{2}\sqrt{\frac{2
c_1}{\gamma}}(x - c t)\right)}{C_1
{\rm{tanh}}\frac{1}{2}\sqrt{\frac{2 c_1}{\gamma}}(x - c t) +
C_2}\right] + \frac{B}{3 C}.
\end{eqnarray}
\\
\\
\textbf{Case 2}:\,\,\,\,$\displaystyle \mu = 0,\,\,\lambda = -
\sqrt{\frac{2 c_1}{\gamma}}.$

\begin{eqnarray}
u_2(x,t) = \displaystyle \pm \sqrt{- \frac{c_1}{c_2}}\left[1 +
\frac{(C_1 + C_2)\left(1 - {\rm{tanh}}\frac{1}{2}\sqrt{\frac{2
c_1}{\gamma}}(x - c t)\right)}{C_2 - C_1
{\rm{tanh}}\frac{1}{2}\sqrt{\frac{2 c_1}{\gamma}}(x - c t)}\right] +
\frac{B}{3 C}.
\end{eqnarray}
\\
\\
\textbf{Case 3}:\,\,\,\,$\displaystyle \lambda = 0,\,\,\mu = -
\frac{c_1}{2 \gamma}.$

\begin{eqnarray}
u_3(x,t) = \displaystyle \pm \sqrt{- \frac{c_1}{c_2}}\left[\frac{C_1
+ C_2 {\rm{tanh}}\frac{1}{2}\sqrt{\frac{2 c_1}{\gamma}}(x - c
t)}{C_1 {\rm{tanh}}\frac{1}{2}\sqrt{\frac{2 c_1}{\gamma}}(x - c t) +
C_2}\right] + \frac{B}{3 C}.
\end{eqnarray}
In all three cases,  $\gamma$ and $c_1$ should have the same signs
and $c_2$ should be of opposite sign and  $C_1 \neq \pm C_2$.
\\
\\
Figure 1 and Figure 2 represent the solutions given by equation
(29).
\begin{figure}[h!]
\begin{center}
\includegraphics[width=3in]{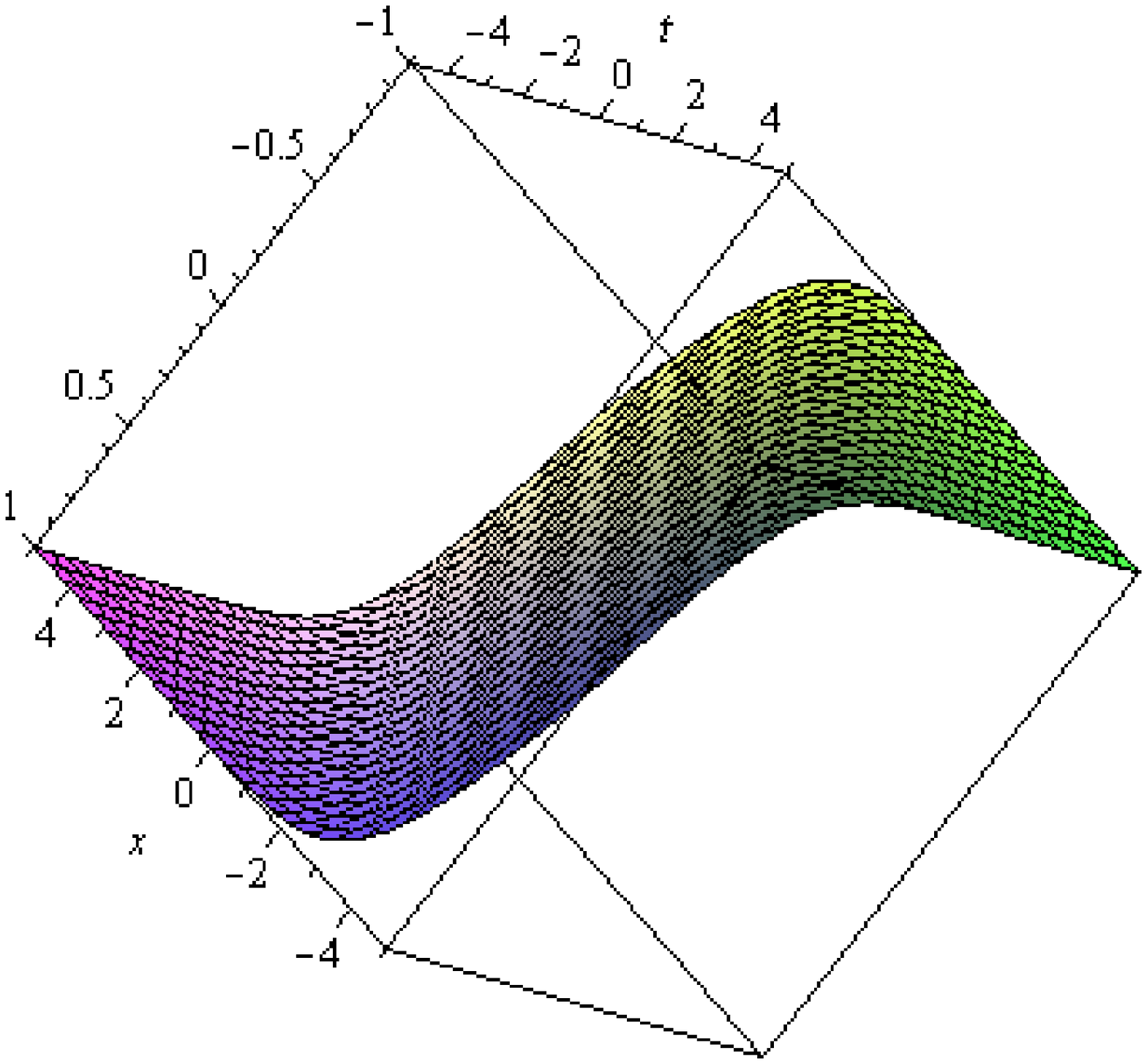}
\caption{\small The solution for $u(x,t)$, $C_{1}=0,C_{2}=1$.}
\end{center}
\end{figure}

\begin{figure}[h!]
\begin{center}
\includegraphics[width=3in]{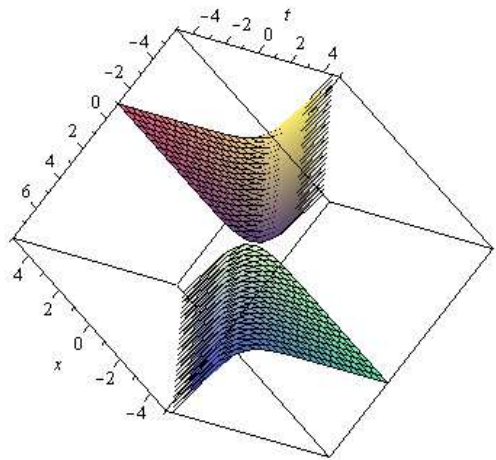}
\caption{\small The solution for $u(x,t)$, $C_{1}=1,C_{2}=0$.}
\end{center}
\end{figure}
Using equation (12), we can write down the corresponding solutions
$v_1(x,t),\,v_2(x,t)$ and $v_3(x,t)$.

\section{Weierstrass Elliptic Function Solutions of the Coupled Wave Equation}

The Weierstrass elliptic function (WEF) $\wp(\xi;g_2,g_3)$ with
invariants $g_2$ and $g_3$ satisfy
\\
\begin{eqnarray}
\displaystyle {\wp^{\prime}}^2 = 4 \wp^3 - g_2 \wp - g_3,
\end{eqnarray}
\\
where $g_2$ and $g_3$ are related by the inequality
\\
\begin{eqnarray}
\displaystyle g_2^3 - 27 g_3^2 > 0.
\end{eqnarray}
\\
The WEF $\wp(\xi)$ is related to the JEFs by the following
relations:
\\
\begin{eqnarray}
\displaystyle {\rm{sn}}(\xi) = \left[\wp(\xi)-e_3\right]^{-1/2},
\end{eqnarray}
\begin{eqnarray}
\displaystyle {\rm{cn}}(\xi) =
\left[\frac{\wp(\xi)-e_1}{\wp(\xi)-e_3}\right]^{1/2},
\end{eqnarray}
\begin{eqnarray}
\displaystyle {\rm{dn}}(\xi) =
\left[\frac{\wp(\xi)-e_2}{\wp(\xi)-e_3}\right]^{1/2},
\end{eqnarray}
\\
where $e_1,e_2,e_3$ satisfy
\\
\begin{eqnarray}
4z^3 - g_2 z - g_3 = 0
\end{eqnarray}
\\
with
\\
\begin{eqnarray}
\displaystyle e_1 = \frac{1}{3}(2 - m^2),\,\,\,e_2 =
\frac{1}{3}(2m^2 - 1),\,\,\,e_3 = -\frac{1}{3}(1+m^2).
\end{eqnarray}
\\
From equation (38), one can see that the modulus $m$ of the JEF and
the $e's$ of the WEF are related by
\\
\begin{eqnarray}
\displaystyle m^2 = \frac{e_2 - e_3}{e_1 - e_3}.
\end{eqnarray}
\\
We consider the ODE of order $2k$ given by
\begin{eqnarray}
\displaystyle \frac{d^{2k}\phi}{d\xi^{2k}} = f(\phi;r+1),
\end{eqnarray}
\\
where $f(\phi;r+1)$ is an $(r+1)$ degree polynomial in $\phi$.
\\
\\
We assume that
\\
\begin{eqnarray}
\phi = \gamma Q^{2s}(\xi) + \mu
\end{eqnarray}
\\
is a solution of equation (40), where $\gamma$ and $\mu$ are
arbitrary constants and $Q^{(2s)}(\xi)$ is the $(2s)^{\rm{th}}$
derivative of the reciprocal Weierstrass elliptic function (RWEF)
$\displaystyle Q(\xi) = \frac{1}{\wp(\xi)}, \wp(\xi)$ being the WEF.
\\
\\
It can be shown that the $(2s)^{\rm{th}}$ derivative of the RWEF
$Q(\xi)$ is a $(2s+1)$ degree polynomial in $Q(\xi)$ itself.
\\
\\
Therefore, for $\phi$ to be a solution of equation (40), we should
have the relation
\\
\begin{eqnarray}
2k - r = 2rs.
\end{eqnarray}
\\
So, it is necessary that $2k \geq r$ for us to assume a solution in
the form of equation (41). But this is in no way a sufficient
condition for the existence of the PWS in the form of equation (41).
\\
\\
Now, we shall search for the WEF solutions of equation (17). We
introduce a restriction on the coefficients in the form $2B^2 =
9AC$, so that equation (17) reduces to
\begin{equation}
\gamma w^{\prime\prime} + c_1 w + c_2 w^3  = 0,
\end{equation}
where $\displaystyle c_1 = -\frac{1}{2}\,A,\,\,c_2 = C$.
\\
\\
For a solution of equation (43) in the form of equation (41), we
should have $r = 2$ and $k = 1$ so that $s = 0.$ So, our solution
will be
\begin{eqnarray}
u(\xi) = \displaystyle \frac{\tau}{\wp(\xi)} + \zeta.
\end{eqnarray}

Substituting equation (44) into equation (43) and equating the
coefficients of like powers of $\wp(\xi)$ to zero, we obtain

\begin{eqnarray}
\wp^3(\xi):\,\,\,\,\displaystyle 2 \gamma \tau - \frac{1}{2} A \zeta
+ C \zeta^3 = 0,
\end{eqnarray}
\begin{eqnarray}
\wp^2(\xi):\,\,\,\,\displaystyle  - \frac{1}{2} A \tau + 3 C \tau
\zeta^2 = 0,
\end{eqnarray}
\begin{eqnarray}
\wp(\xi):\,\,\,\,\displaystyle - \frac{3}{2} \gamma \tau g_2 + 3 C
\tau^2 \zeta = 0,
\end{eqnarray}
\begin{eqnarray}
\wp^0(\xi):\,\,\,\,\displaystyle  - 2 \gamma \tau g_3 + C \tau^3 =
0.
\end{eqnarray}

From equations (46) through (48), it can be found that

\begin{eqnarray}
\displaystyle \tau = \pm \sqrt{\frac{2 \gamma g_3}{C}},
\end{eqnarray}
\begin{eqnarray}
\displaystyle \zeta = \pm \sqrt{\frac{A}{6C}},
\end{eqnarray}
\begin{eqnarray}
\displaystyle g_2 = 2 \sqrt{\frac{A g_3}{3 \gamma}}.
\end{eqnarray}

From equations (49) through (51), one can conclude that if $g_3 >
0$, $A,\,C$ and $\gamma$ should all be of the same signs, whereas
for $g_3 < 0$, $A$ and $C$ should be of the same signs and $\gamma$
should be of opposite sign.
\\
\\
Equation (45) leads us to the value of $g_3$ given by

\begin{eqnarray}
\displaystyle g_3 = \frac{A^3}{432 \gamma^3}.
\end{eqnarray}

The condition $\displaystyle g_2^3 - 27 g_3^2 > 0$ gives the
relation

\begin{eqnarray}
\displaystyle  \frac{8}{9} > \frac{3}{4},
\end{eqnarray}

which is remarkably true for any value of the coefficients in the
governing equation.
\\
\\
The equations (34) through (36) will give rise to the same PWS of
equation (43) which can be obtained using equation (44) with the
help of equation (38). Thus the PWS of equation (43) in terms of
JEFs can be written as
\begin{eqnarray}
u(\xi) = \displaystyle \frac{\tau \, {\rm{sn}}^2(\xi)}{1-\frac{1}{3}
(1+m^2){\rm{sn}}^2(\xi)}+ \zeta.
\end{eqnarray}

As $m \rightarrow 1$, the SWSs of the coupled wave equation given by
equations (6) and (7) with the restriction $2B^2 = 9AC$ are

\begin{eqnarray}
u(x,t) = \displaystyle \frac{B}{3C} + \frac{\tau \,
{\rm{tanh}}^2(x-ct)}{1-\frac{2}{3}{\rm{tanh}}^2(x-ct)}+ \zeta,
\end{eqnarray}
\begin{eqnarray}
v(x,t) = \displaystyle \frac{2c}{\epsilon} -
\frac{2\sigma}{\epsilon}\left[\frac{B}{3C} + \frac{\tau \,
{\rm{tanh}}^2(x-ct)}{1-\frac{2}{3}{\rm{tanh}}^2(x-ct)}+
\zeta\right],
\end{eqnarray}

where $\tau$ and $\zeta$ are given by eqs. (49) and (50).

\section{Conclusions}
The $(G'/G)$-expansion method has been applied to a nonlinear
coupled wave equation. The kink wave solution and the singular wave
solution have been graphically illustrated. It was found that there
are some restrictions on the coefficients in the governing equation
for the solutions in terms of hyperbolic functions to exist. The WEF
method has also been applied to the system of equations to derive
SWSs. The condition $\displaystyle g_2^3 - 27 g_3^2 > 0$ was found
to be identically satisfied, which is a remarkable result and has
never been reported in the literature. We intend to apply these
methods for higher order and higher dimensional PDEs of physical
interest.


\begin{thebibliography}{99}


\bibitem{R1} A. Biswas. Solitary wave solution for KdV equation with power
 law nonlinearity and time-dependent coefficients.{\em Nonlinear Dynamics} {\bf{58}}
(2009) 345-348.

\bibitem{R2} A. Biswas. Solitary waves for power-law regularized long wave equation
 and R(m,n) equation. {\em  Nonlinear Dynamics} {\bf{59}} (2010) 423-426.

\bibitem{R3} R. Sassaman and A. Biswas. Topological and  non-topological solitons of the
 Klein-Gordon equations in $(1+2)$-dimensions. {\em  Nonlinear Dynamics} {\bf{61}} (2010) 23-28.

\bibitem{R4} A. Biswas, A.H. Kara, A.H. Bokhari and F.D. Zaman. Solitons and conservation laws of Klein-Gordon
 equation with power law and log law nonlinearities. {\em  Nonlinear Dynamics} {\bf{73}} (2013) 2191-2196.

\bibitem{R5} M. Mirzazadeh, M. Eslami, E. Zerrad, M.F. Mahmood, A. Biswas and M. Belic. Optical solitons in nonlinear
 directional couplers by sine-cosine function method and Bernoulli's equation approach.{\em Nonlinear Dynamics} {\bf{81}} (2015) 1933-1949.

\bibitem{R6} W. Malfliet. The tanh method: Exact solutions of nonlinear evolution and wave equations. {\em  Physica Scripta} {\bf{54}} (1996) 563-568.

\bibitem{R7} M. Alquran and K. Al-Khaled K., Sinc and Solitary Wave Solutions to the Generalized
Benjamin-Bona-Mahony-Burgers Equations. {\em Physica Scripta}
{\bf{83}} (2011) 065010 (6 pp).

\bibitem{R8} M. Alquran and K. Al-Khaled. The tanh and sine-cosine methods for higher order
 equations of Korteweg-de Vries type. {\em Physica Scripta} {\bf{84}} (2011) 025010 (4 pp).

\bibitem{R9} S. Shukri and K. Al-Khaled. The Extended Tanh Method For Solving Systems of
Nonlinear Wave Equations. {\em Applied Mathematics and Computation}
{\bf{217}} (2010) 1997-2006.

\bibitem{MAR5} M. Alquran, H.M. Jaradat and M. Syam. A modified approach for a reliable
 study of new nonlinear equation: two-mode Korteweg-de Vries-Burgers equation.
\emph{Nonlinear Dynamics} \textbf{91}(3) (2018) 1619-1626.

\bibitem{MAR6} A. Jaradat, M.S.M. Noorani, M. Alquran and H.M. Jaradat. A Variety of New Solitary-Solutions for the Two-mode
Modified Korteweg-de Vries Equation. \emph{Nonlinear Dynamics and
Systems Theory} \textbf{19}(1) (2019) 88-96.

\bibitem{R10} J.H. He and X.H. Wu. Exp-function method for nonlinear wave equations. {\em Chaos Solitons and Fractals}
{\bf{30}} (2006) 700-708.

\bibitem{R11} J. Liu, L. Yang and K. Yang. Jacobi elliptic function solutions of some
 nonlinear PDEs. {\em Physics Letters A} {\bf{325}} (2004) 268-275.

\bibitem{MAR1} M. Alquran and A. Jarrah. Jacobi elliptic function solutions for a two-mode KdV equation. \emph{Journal of King Saud
University-Science} (2017).
https://doi.org/10.1016/j.jksus.2017.06.010.

\bibitem{MAR2} M. Alquran, A. Jarrah and E.V. Krishnan. Solitary wave solutions of the phi-four equation and the breaking soliton system
by means of Jacobi elliptic sine-cosine expansion method.
\emph{Nonlinear Dynamics and Systems Theory} \textbf{18}(3) (2018)
233-240.

\bibitem{extra} M. Al Ghabshi, E.V. Krishnan and M. Alquran. Exact solutions of a Klein-Gordon system by
$(G'/G)$-expansion method and Weierstrass elliptic function method.
\emph{Nonlinear Dynamics and Systems Theory} \textbf{19}(3) (2019)
386-395.

\bibitem{R12} Y. Peng. Exact periodic wave solutions to a new Hamiltonian amplitude equation. {\em J of the Phys Soc of Japan}, {\bf{72}} (2003) 1356-1359.

\bibitem{R13} Y. Peng. New Exact solutions to a new Hamiltonian amplitude equation. {\em J of the Phys Soc of Japan}
{\bf{72}} (2003) 1889-1890.

\bibitem{R14} Y. Peng. New Exact solutions to a new Hamiltonian amplitude equation II. {\em J of the Phys Soc of Japan} {\bf{73}} (2004) 1156-1158.

\bibitem{R15} E.V. Krishnan and Y. Peng. A new solitary wave solution for the new Hamiltonian
 amplitude equation. {\em Journal of the Physical Society of Japan} {\bf{74}} (2005) 896-897.


\bibitem{R16} J.F. Alzaidy. Extended mapping method and its applications to nonlinear evolution
 equations. {\em Journal of Applied Mathematics} {\bf{2012(2012)}} Article ID 597983, 14 pages.

\bibitem{R17} M. Al Ghabshi, E.V. Krishnan, K. Al-Khaled and M. Alquran. Exact and Approximate Solutions of a System of Partial
 Differential Equations, {\em International Journal of Nonlinear Science}. {\bf{23}} (2017) 11-21.

\bibitem{jaradat1} M. Syam, H.M. Jaradat and M. Alquran. A study on the two-mode
coupled modified Korteweg-de Vries using the simplified bilinear and
the trigonometric-function methods. \emph{Nonlinear Dynamics}
\textbf{90}(2)(2017) 1363-1371.

\bibitem{jaradat2} H.M. Jaradat, M. Syam and M. Alquran. A two-mode coupled
Korteweg-de Vries: multiple-soliton solutions and other exact
solutions. \emph{Nonlinear Dynamics} \textbf{90}(1)(2017) 371-377.


\bibitem{trig1} M. Alquran. Bright and dark soliton solutions to the
Ostrovsky-Benjamin-Bona-Mahony (OS-BBM) equation. \emph{Journal of
Mathematical and Computational Science} \textbf{2}(1) (2012) 15-22.

\bibitem{trig2} M. Alquran, R. Al-Omary and Q. Katatbeh. New explicit solutions for homogeneous KdV equations of third order
by trigonometric and hyperbolic function methods. \emph{Applications
and Applied Mathematics} \textbf{7}(1) (2012) 211-225.

\bibitem{trig3} M. Alquran. Solitons and periodic solutions to nonlinear partial differential
equations by the sine-cosine method. \emph{Applied Mathematics and
Information Sciences} \textbf{6}(1) (2012) 85-88.


\bibitem{R18} R. Hirota and J. Satsuma. Soliton solutions of a coupled Korteweg-de Vries equation. {\em  Physics Letters A} {\bf{85}} (1981) 407-408.

\bibitem{R19} R. Dodd and A.P. Fordy. On the integrability of a system of coupled KdV equations. {\em Physics Letters A} {\bf{89}} (1982) 168-170.

\bibitem{R20} H. Gao and  R.X. Zhao. New application of the $(G'/G)$-expansion method to high-order nonlinear equations. {\em Appl Math Computation}
{\bf{215}} (2009) 2781-2786.

\bibitem{R21} F. Chand F. and  A.K. Malik. Exact traveling wave solutions of some nonlinear equations
 using $(G'/G)$-expansion method. {\em International Journal of Nonlinear Science} {\bf{14}} (2012) 416-424.

\bibitem{R22}  M. Alquran and  A. Qawasmeh. Soliton solutions of shallow water wave
 equations by means of  $(G'/G)$-expansion method. {\em Journal of Applied Analysis and Computation} {\bf{4}} (2014) 221-229.

\bibitem{MAR3} O. Yassin and M. Alquran. Constructing New Solutions for Some
Types of Two-Mode Nonlinear Equations. \emph{Applied Mathematics and
Information Sciences} \textbf{12}(2) (2018) 361-367.

\bibitem{MAR4} M. Alquran and O. Yassin. Dynamism of two-mode's parameters on the field
function for third-order dispersive Fisher: Application for fibre
optics. \emph{Optical and Quantum Electronics} \textbf{50}(9) (2018)
354.


\bibitem{R23} E.V. Krishnan. Remarks on a system of coupled nonlinear wave equations. {\em Journal of Mathematical Physics}, {\bf{31}} (1990) 1155-1156.

\bibitem{R24} D.W. Lawden. Elliptic functions and applications. Springer Verlag, Berlin (1989).


\end{thebibliography}
\end{document}